\documentstyle[12pt,worldsci]{article}
\input epsf
\newenvironment{indention}[1]{\par
\addtolength{\leftskip}{#1}
\begingroup}{\endgroup\par}

\begin{document}
 \title{Intersection between Microscopic and Macroscopic Abelian Dominance
\\ in the Confinement Physics of QCD}
\author{Hiroko Ichie\footnote{ E-mail : ichie@th.phys.titech.ac.jp} \\
{\em Department of Physics, Tokyo Institute of Technology, 
Tokyo, 152-8551, Japan}\\
\vspace{0.3cm}
Hideo Suganuma\\
{\em Research Center for Nuclear Physics, 
Osaka University, 
 Osaka, 567-0047, Japan}}
\maketitle
\setlength{\baselineskip}{2.6ex}

\vspace{0.3cm}
\begin{abstract}
We study abelian dominance for confinement
in terms of the local gluon properties
in the maximally abelian (MA) gauge,
where the diagonal component of the gluon 
is maximized by the gauge transformation.
We find microscopic abelian dominance on the link-variable
for the whole region of $\beta$ in the lattice QCD in the MA gauge.
The off-diagonal angle variable, which is not constrained
by the MA-gauge fixing condition,
tends to be random besides the residual gauge degrees of freedom.
Within the random-variable approximation for
the off-diagonal angle variable,
we analytically prove that off-diagonal gluon contribution $W^{\rm off}$
to the Wilson loop
obeys the perimeter law in the MA gauge.
The perimeter-law behavior of $W^{\rm off}$ is also confirmed 
using the lattice QCD simulation.
This indicates macroscopic abelian dominance for the string tension.
\end{abstract}
\vspace{0.3cm}

\section{Introduction}

In the low-energy region of QCD, there appear interesting 
phenomena such as color confinement and chiral symmetry breaking  reflecting
the strong gauge-coupling.
However, 
because of nonperturbative and nonabelian nature,
these phenomena are difficult to treat analytically,
and 
it is desired to extract the relevant degrees of freedom 
 for description of infrared phenomena.

In 1974, Nambu 
proposed an 
idea that quark confinement can be interpreted 
using the dual version of the 
superconductivity\cite{nambu}. 
In the 
superconductor, 
Cooper-pair condensation leads to the Meissner effect, 
and the magnetic flux is 
 squeezed like a 
quasi-one-dimensional tube as the Abrikosov vortex.
In this dual-superconductor picture for the QCD vacuum, 
the squeezing of the color-electric flux between quarks 
is realized by the dual Meissner effect
as the result of condensation of color-magnetic monopoles.
However, there are two following large gaps between QCD and the dual 
superconductor picture;
1) This picture is based on the abelian gauge theory, 
while QCD is a nonabelian gauge theory. 
2) The dual-superconductor scenario requires condensation of magnetic 
monopoles as the key concept, while QCD does not have such a monopole as 
the elementary degrees of freedom. 
As the connection between QCD and the dual superconductor scenario, 
't Hooft proposed the concept of the abelian gauge fixing\cite{thooft}
with assumption of abelian dominance for the infrared QCD. 
The abelian gauge fixing is defined so as to diagonalize a suitable 
gauge-dependent variable $\Phi[A_\mu(x)]$
and
reduces QCD into an abelian gauge
theory, where the off-diagonal element of the gluon field behaves as a 
charged matter field. 
Moreover, in the abelian gauge, color-magnetic monopoles appear
as topological objects corresponding to the
nontrivial homotopy group $\Pi_2( {\rm SU(N_c)/U(1)}^{\rm N_c-1}) =
{\bf Z}^{\rm N_c-1}_\infty$.
If monopole condenses, 
the scenario of color confinement by the dual Meissner effect would be realized
in QCD.
In this paper, with the help of the lattice QCD simulation,
we study intersection between abelian dominance of the gluon field 
(microscopic variable) and confinement force
(macroscopic variable) as the theoretical basis of dual superconductor 
picture.

\section{Microscopic Abelian Dominance in the Maximally Abelian Gauge}

Abelian dominance on the confinement force 
have been investigated using 
the lattice QCD simulation in the maximally abelian (MA)
gauge\cite{ichiead}.
In terms of the link variable $U_\mu(s) \equiv U_\mu^0(s) + i\tau^a 
U_\mu^a(s)$, the MA gauge fixing is defined by maximizing 
$R \equiv \sum_{s,\mu} {\rm tr}\{  U_\mu(s) \tau_3 U^{\dagger}_\mu(s) \tau_3 \}
= \sum_{s,\mu}\{ 
( U^0_\mu(s))^2+(U^3_\mu(s))^2-(U^1_\mu(s))^2-(U^2_\mu(s))^2 \}$
through the gauge transformation.
In the MA gauge, the off-diagonal components, $U_\mu^1$ and $U_\mu^2$, are 
forced to be small,
and therefore the QCD system seems describable only by U(1)-like variables 
approximately.
The MA gauge is a sort of the abelian gauge, because
the MA gauge fixing diagonalizes 
$\Phi(s) \equiv \sum _{\mu,\pm } U_{\pm \mu}(s) \tau_3 U^{\dagger}_{\pm \mu}(s)$
with $U_{-\mu}(s) \equiv 
U^{\dagger}_\mu(s-\mu)$.
In this section, we study 
abelian dominance on the link variable
$U_{\mu}(s)$.

In the lattice formalism, the SU(2) link variable 
$U_\mu$(s) is
factorized as 
%
\begin{eqnarray}
{\small
 U_\mu(s) 
=\left( {\matrix{
{\rm cos}{\theta_\mu}(s) & -{\rm sin}{\theta_\mu}(s) e^{-i\chi_\mu(s)} \cr
{\rm sin}{\theta_\mu}(s) e^{i\chi_\mu(s)} & {\rm cos}{\theta_\mu}(s)
}} \right)
\left( \begin{array}{cc} e^{i\theta^3_\mu(s)} & 0 \\ 0 & e^{-i\theta^3_\mu(s)}
 \end{array} \right) 
\equiv M_\mu (s) u_\mu(s). \nonumber
}
\end{eqnarray} 
Here, the U(1)$_{3}$ link-variable $u_\mu(s)$ corresponds to the diagonal gluon 
part
and behaves as the 
abelian gauge filed in the MA gauge, while $M_\mu(s)$
corresponds to the off-diagonal gluon part.

In order to investigate  abelian dominance on the link variable in the 
MA gauge,
we define ``abelian projection rate'' \cite{ichiead}  as $R_{\rm Abel} = \cos 
\theta_\mu(s)$ $\in$ $[0,1]$ with 
$0 \le \theta_{\mu} \le \frac{\pi}{2}$. 
For instance, 
the SU(2) link variable becomes completely diagonal 
if $\cos \theta  =1$, 
while  it becomes 
off-diagonal
if  $\cos \theta =0$. 
In Fig.1, we show local abelian projection rate
$R_{\rm Abel}$  
expressed by the  arrow $(\sin \theta, \cos \theta)$ in a 
typical configuration of the lattice QCD.
In the MA gauge, most of all SU(2) link variables become U(1)-like.
For the quantitative argument, we show in Fig.2 the probability distribution  
$P(R_{\rm Abel})$  
of the abelian projection rate $R_{\rm Abel}$. 
Without gauge fixing, one finds the 
average 
$\langle  R_{\rm Abel} \rangle = \frac23 $. 
In the MA 
gauge, the off-diagonal
component of the SU(2) link variable 
is forced to be reduced, and
$R_{\rm Abel}$ approaches to unity;
one obtains   
$\langle  R_{\rm Abel} \rangle_{\rm MA} \simeq$  0.93
on $16^4$ lattice
with $\beta = 2.4$.
Thus, we find  {\it microscopic abelian dominance} on the link variable.



\newlength{\miniwocolumn}

\setlength{\miniwocolumn}{0.50\textwidth}

\begin{indention}{-1.2cm}
\begin{minipage}[t]{\miniwocolumn}
\vspace{0.2cm}
\epsfxsize = 8.8 cm
\epsfbox{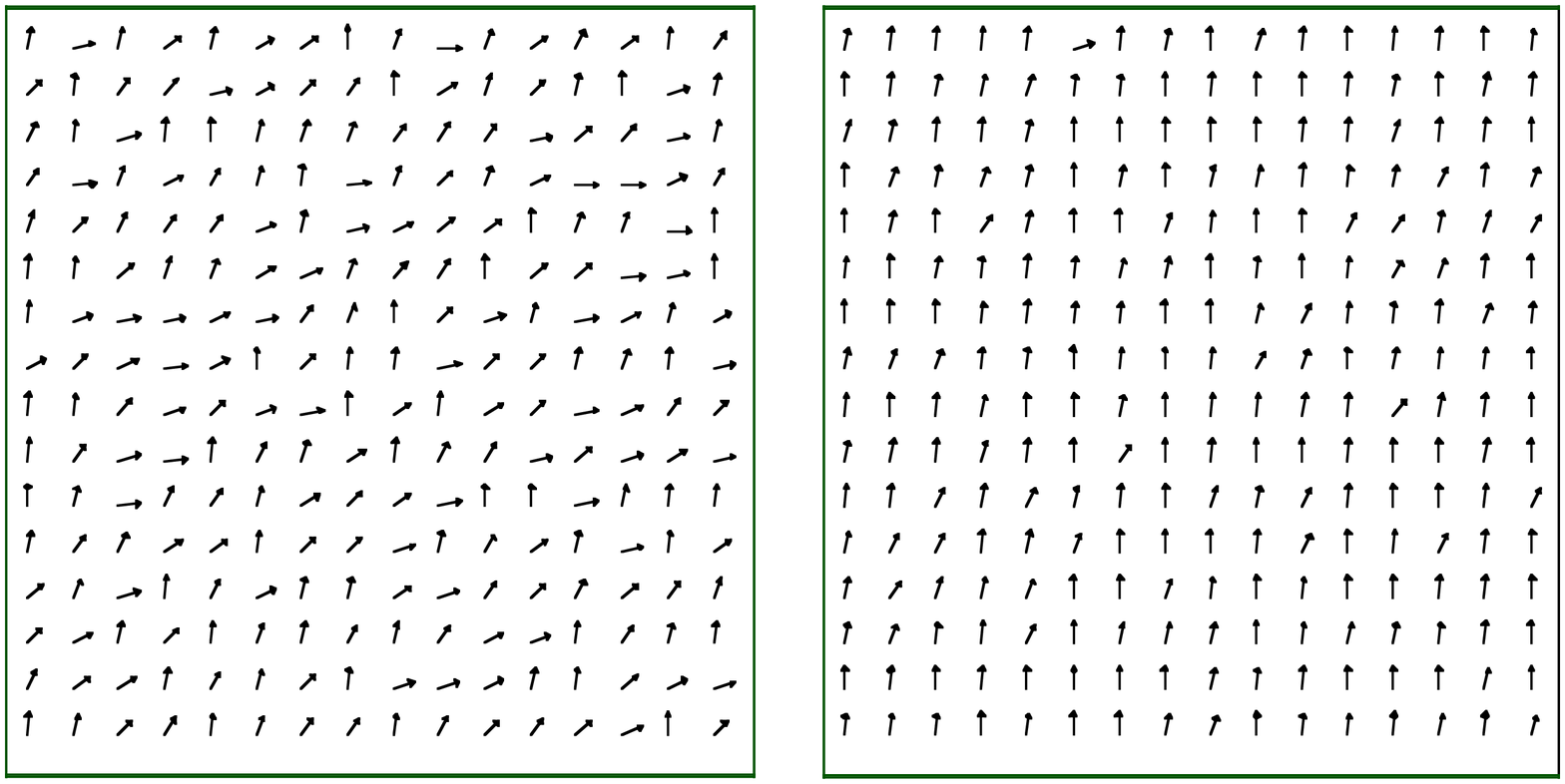}
\vspace{-2.4cm}

{\footnotesize 
{\bf Fig.1} Local abelian projection rate 
$R_{\rm Abel}$ $\equiv$ $\cos \theta $ 
$(0 \le \theta_{\mu} \le \frac{\pi}{2})$  
at $\beta$=  2.4  on  
$16^4$ lattice without gauge fixing (left) and in
MA gauge fixing (right). 
The arrow expresses  $(\sin \theta, \cos \theta).$
}
\label{arrow5}
\end{minipage}
\hspace{0.4cm}
\setlength{\miniwocolumn}{0.45\textwidth}
\begin{minipage}[t]{\miniwocolumn}
\vspace{0.2cm}
\epsfxsize = 6.0cm
\centerline{\epsfbox{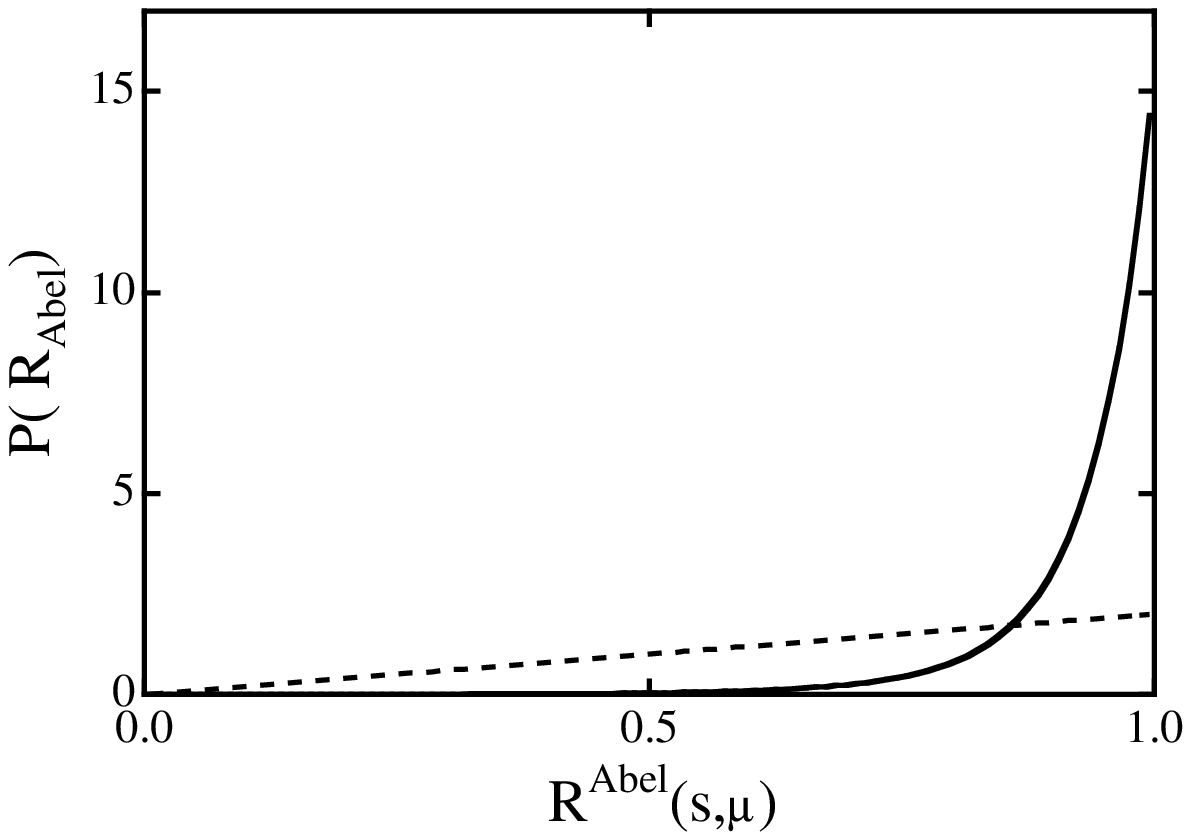}}

{\footnotesize {\bf Fig.2}
 The probability distribution $P(R_{\rm Abel})$ 
of abelian projection rate $R_{\rm Abel} $
at $\beta =2.4$ on $16^4$ lattice 
in the MA gauge (solid curve) and without gauge fixing (dashed curve).
}
\label{arrow6}
\end{minipage}
\vspace{0.2cm}
\end{indention}

\vspace{0.2cm}

\noindent

\section{Semi-analytical Proof of 
Abelian Dominance for Confinement} 


\indent 
In the MA gauge, the {\rm diagonal element} $\cos \theta_\mu(s)$ 
in $M_\mu(s)$ is maximized by the gauge transformation. 
Then, the  off-diagonal element 
$e^{i\chi_\mu(s)}\sin\theta_\mu(s)$ is forced to take a small value 
in the MA gauge, and therefore the approximate treatment 
on the off-diagonal element would be allowed in the MA gauge.
Moreover, the angle variable $\chi_\mu(s)$ is not 
constrained by the MA gauge-fixing condition at all, 
and  tends to take a random value\cite{ichiead} besides the residual 
${\rm U(1)}_3$ gauge degrees 
of freedom.
Hence, 
$\chi_\mu(s)$ in the MA gauge can be regarded 
as a {\it random angle variable} 
in a good approximation. 
In this section, we first investigate properties of $\chi_\mu(s)$
using the lattice QCD simulation, and then
study the origin of abelian dominance on confinement. 

We examine the randomness of $\chi_\mu(s)$
using the lattice QCD simulation in the MA gauge with 
U(1)$_3$ Landau-gauge fixing\cite{ichiead}.
We show in Fig.3
the probability distribution $P(\Delta \chi)$
of the correlation
$\Delta \chi(s) \equiv 
{\rm mod}_{\pi}|\chi_\mu(s)-\chi_\mu(s+ \hat \nu)| \in [0,\pi]$,
which is the difference between two neighboring angle variables,
at $\beta$=0, 1.0, 2.4, 3.0.
In the strong-region  as $\beta \le 1$,
$\chi_\mu(s)$ behaves as a random variable, and
there is no correlation between neighboring $\chi_\mu$.
On the other hand, 
in the weak-coupling region,
the smallness of $\sin \theta_\mu$ makes off-diagonal components
more irrelevant in the MA gauge, which permits 
the approximate treatment 
on $\chi_\mu(s)$.
Thus, we can take the random-variable approximation for $\chi_\mu(s)$
as a good approximation
in the whole region of $\beta$ in the MA gauge.


Next, let us consider the Wilson loop 
$\langle W_C[U_\mu(s)]\rangle \equiv 
\langle{\rm tr}\Pi_C U_\mu(s)\rangle$
in the MA gauge.
In calculating $\langle W_C[U_\mu(s)]\rangle$, 
the expectation value of $e^{i\chi_\mu(s)}$
in $M_\mu(s)$ vanishes as
$\langle e^{i\chi_\mu(s)}\rangle 
\simeq \int_0^{2\pi} d\chi_\mu(s)\exp\{i\chi_\mu(s)\}=0$
within the random-variable approximation on $\chi_\mu(s)$.
Then, the  off-diagonal factor 
$M_\mu(s)$ appearing 
in the Wilson loop  $W_C[U_\mu(s)]$ 
becomes a diagonal matrix,
$U_\mu(s)\equiv M_\mu(s)u_\mu(s)
\rightarrow 
\cos \theta_\mu(s) u_\mu(s).
$

Then, for the $I \times J$ rectangular $C$, the Wilson loop 
{\small 
$W_C[U_\mu(s)]$  $\equiv$ $\langle{\rm tr}\Pi_{i=1}^L U_{\mu_i}(s_i)\rangle$
}
in the MA gauge is estimated as 
{\small
\begin{eqnarray}
\langle W_C[U_\mu(s)]\rangle
& \simeq & 
\langle {\rm tr}\Pi_{i=1}^L \cos \theta_{\mu_i}(s_i) u_{\mu_i}(s_i)
\rangle_{\rm MA} =
\langle \Pi_{i=1}^L \cos \theta_{\mu_i}(s_i) \cdot 
{\rm tr} \Pi_{j=1}^L u_{\mu_j}(s_j)\rangle_{\rm MA} \nonumber \\
&\simeq& 
\exp\{L \langle \ln (\cos \theta_\mu(s)) \rangle_{\rm MA} \} 
\ \langle W_C[u_\mu(s)]\rangle_{\rm MA},
\end{eqnarray}
}
\noindent
where $L\equiv 2(I+J)$ denotes the perimeter length and 
$W_C[u_\mu(s)]\equiv {\rm tr}\Pi_{i=1}^L u_{\mu_i}(s_i)$ 
the abelian Wilson loop.
Here, we have replaced 
$\sum_{i=1}^L \ln \{\cos(\theta_{\mu_i}(s_i)\}$ 
by its average 
$L \langle \ln \{\cos \theta_\mu(s)\} \rangle_{\rm MA}$
 in a statistical sense.
In this way, we derive a simple estimation as 
\begin{eqnarray}
W_C^{\rm off}\equiv 
\langle W_C[U_\mu(s)]\rangle/\langle W_C[u_\mu(s)]\rangle_{\rm MA}
\simeq \exp\{L\langle \ln(\cos \theta_\mu(s))\rangle_{\rm MA}\}
\end{eqnarray}
for the contribution of the off-diagonal 
gluon element to the Wilson loop.
From this analysis, 
$W_C^{\rm off}$ is expected to obey the {\it perimeter law} 
in the MA gauge for large loops, where the statistical 
treatment would be accurate.

In the lattice QCD, we find that 
$W_C^{\rm off}$ seems to obey the 
 perimeter law for the Wilson loop with $I,J \ge 2$ 
in the MA gauge (Fig.4). 
We find also that the lattice data of $W_C^{\rm off}$ 
as the function of $L$ are well reproduced 
by the above analytical estimation with   microscopic information
on 
$\cos\theta_\mu(s)$ as 
$\langle \ln \{\cos \theta_\mu(s)\} \rangle_{\rm MA}\simeq -0.082$ 
for $\beta=2.4$.


Thus, the off-diagonal contribution $W_C^{\rm off}$ 
to the Wilson loop obeys 
the perimeter law in the MA gauge, and therefore 
the SU(2) string-tension 
$\sigma_{\rm SU(2)}$
$\equiv$ 
$-\lim_{I,J \rightarrow \infty}
{1 \over IJ}\ln \langle $ $ W_{I \times J}[U_\mu(s)]\rangle 
$
coincides with to 
the abelian string-tension $\sigma_{\rm Abel}$,
{\small
\begin{eqnarray}
\sigma_{\rm SU(2)}
= -2 \langle \ln \{\cos\theta_\mu(s)\} \rangle_{\rm MA}
{I+J \over IJ}
+ \sigma_{\rm Abel}
\  \ \stackrel{{\small I,J \rightarrow \infty}}{\longrightarrow} \ \
\sigma_{\rm Abel}.
\end{eqnarray}
}
Thus,  abelian dominance for the string tension, 
$\sigma_{\rm SU(2)}=\sigma_{\rm Abel}$, 
can be proved in the MA gauge by 
approximating the off-diagonal angle variable $\chi_\mu(s)$
as a random variable.

\begin{indention}{-0.6cm}
\setlength{\miniwocolumn}{0.47\textwidth}

\begin{minipage}[t]{\miniwocolumn}
\vspace{.1cm}
\epsfxsize = 6.7 cm
\epsfbox{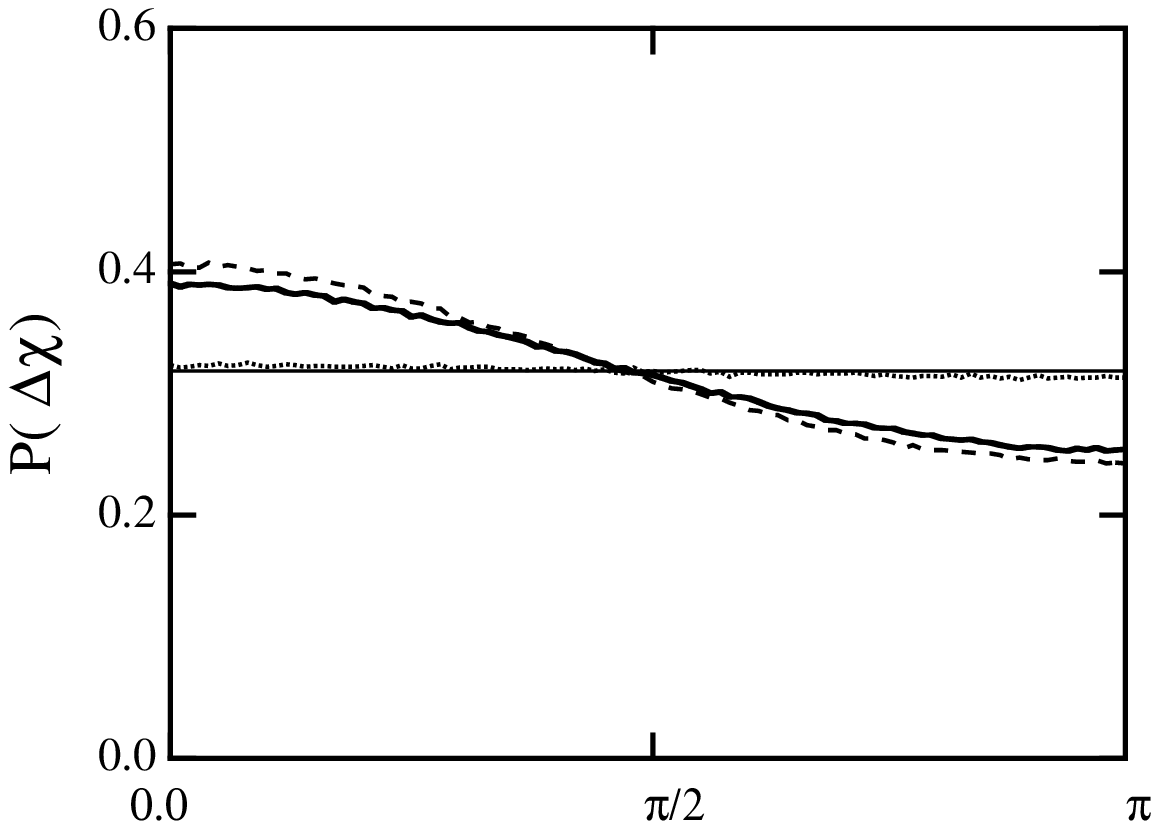}
\vspace{.1cm}
{\footnotesize {\bf Fig.3}
 The probability distribution $P(\Delta \chi)$
of the correlation $\Delta \chi \equiv {\rm 
mod}_{\pi}(|\chi_\mu(s)-\chi_\mu(s+ \hat \nu)|)$ in the MA gauge
with U(1)$_3$ Landau-gauge fixing 
at $\beta$ = 0 (thin line), 1.0 (dotted curve), 2.4 (solid curve), 
3.0 (dashed curve).}
\label{gfig3}
\vspace{0.3cm}
\end{minipage}
\hspace{0.3cm}
\begin{minipage}[t]{\miniwocolumn}
\vspace{.2cm}
          \epsfxsize = 7.0cm
          \centerline{\epsfbox{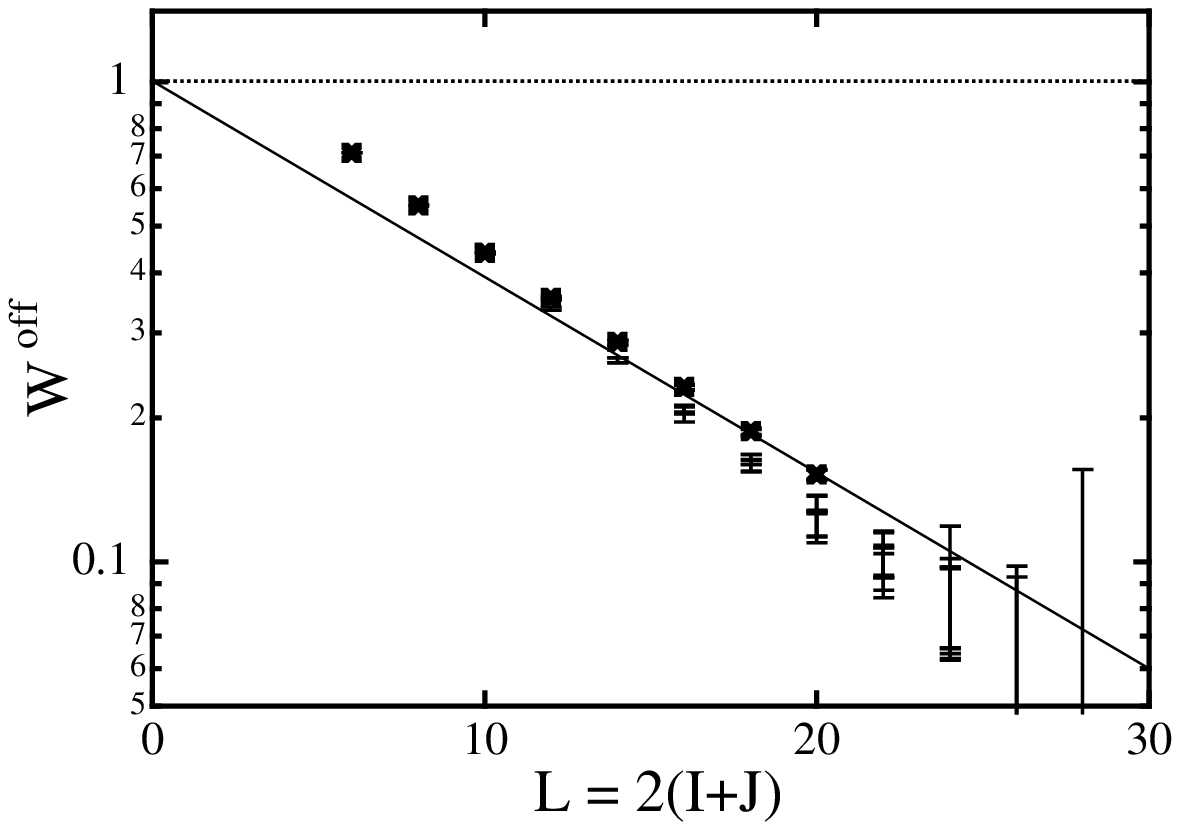}}

{\footnotesize  {\bf Fig.4}
The comparison between 
the analytical estimation (straight line) and
the lattice data ($\times$) of the off-diagonal gluon contribution
$W_C^{\rm off}$
for the Wilson loop  
as the function of 
$L \equiv 2(I+J)$ in the MA gauge
at $\beta=2.4$.
}
                                    \label{fig:off-w}
\end{minipage}
\end{indention}





  
%


\end{document}